\documentclass[10pt,twocolumn,letterpaper]{article}

\usepackage{cvpr}
\usepackage{times}
\usepackage{epsfig}
\usepackage{graphicx}
\usepackage{amsmath}
\usepackage{amssymb}

\def\mH{\mathcal{H}}

\def\mS{\mathcal{S}}
\def\mT{\mathcal{T}}

\def\1n{\mathbf{1}_n}
\def\0{\mathbf{0}}
\def\1{\mathbf{1}}

\def\S{{\bf S}}

\def\a{{\bf a}}
\def\bb{{\bf b}}

\def\bd{{\bf d}}

\def\s{{\bf s}}

\def\y{{\bf y}}

\def\Span{\text{Span}}

\usepackage{amsthm}
\usepackage{caption}
\usepackage{subcaption}

\cvprfinalcopy 

\def\cvprPaperID{****} 

\ifcvprfinal\fi
\begin{document}

\title{On the Uniqueness of Group Sparse Coding,\\ Proof of the Block
  ACS Theorem}

\author{Chen Kong and Simon Lucey\\
The Robotics Institute, Cargenie Mellon University\\
5000 Forbes Ave, Pittsburgh PA., USA\\
{\small chenk@cs.cmu.edu, slucey@cs.cmu.edu}
}

\maketitle

\section{Block ACS Reconstruction Theorem}
\label{sec:bACS}
Denote $A$ as an unknown dictionary, $\a\in\mathbb{R}^{K\alpha\times\beta}$ as a block vector divided into $K$ blocks whose size is $\alpha\times\beta$.
Among $K$ blocks, only $s$ blocks are non-zero, so-called active.
For example, $[1,2,0,0,0,0,0,3,0,0]^T$ can be divided into 5 blocks: $[1,2]^T, [0,0]^T, [0,0]^T, [0,3]^T, [0,0]^T$, and among them only 2 blocks $[1, 2]^T, [0,3]^T$ are active, where $\alpha=2, \beta=1, K = 5, s = 2$.
We call this kind of block vector as $s$-block-sparse vector.

{\it The block ACS theorem}: 
As for the dictionary $A$ satisfying block $2s$-RIP~\cite{eldar2009robust} with RIP constant $\delta < 1$, if there exists another dictionary $B$, such that for any $s$-block-sparse vector $\a$, we can find a $s$-block-sparse $\bb$ satisfying
\begin{equation}
A\a = B\bb,
\label{eq:hyp}
\end{equation}
then it follows that
\begin{equation}
A = B(P_\pi\otimes I_\alpha)D,
\end{equation}
where $P_\pi$ is a permutation matrix and $D$ is a block diagonal matrix, and further the size of each block is $\alpha\times\alpha$.

The block ACS theorem provides the guarantee that given enough samples $y^i = A\a^i$, once one finds a dictionary and coefficients modelling measurements in each samples $[y^1, \dots, y^N]$ well, the ambiguity between recovered dictionary and original dicionary is the mix in the blocks and mix between blocks.

Specially, if $\alpha = 1$, block ACS Theorem degenerates to standard ACS therom~\cite{hillar2011ramsey}.

\section{Proof of Block ACS Theorem}
{Recall in block ACS Theorem, the size of blocks in $\a$ is $\alpha\times\beta$.
Let's first prove the theorem in the case when $\beta = 1$ and once it is proven, the general case $\beta\neq1$ is trivial to handle:
We can split sparse causes $\a$ into $[\a_1,\dots, \a_\beta]$, where $\a_i\in\mathbb{R}^{K\alpha}$ and then the hypotheses $A\a = A[\a_1, \dots, \a_\beta] = B\bb = B[\bb_1, \dots, \bb_\beta]$ equals to $A\a_i = B\bb_i$, which degenerates to the situation where $\beta=1$.

To better understand this theorem and prepare for proof in full generality, let us start from a simple case when $s = 1$.
Denote $e^K_i$ as a $K$-dimensional column vector that has $1$ in its $i$-th coordinate and zeros elsewhere.
Pick up a block vector $\a^{ij} = (e^{K}_i\otimes e^{\alpha}_j),i = 1,\dots,K, j = 1,\dots,\alpha$, saying that its $i$-th coordinate in $j$-th block is 1 and zeros elsewhere.
For some matrix $B$ and $1$-block-sparse $\bb^{ij}$, from hypotheses~\ref{eq:hyp}, it follows that 
\begin{equation}
A\a^{ij} = A(e^K_j\otimes e^\alpha_j) = B\bb^{ij} = (e^K_{\pi(i, j)}\otimes I_\alpha){\bf d}_{ij},
\label{eq:hypS1}
\end{equation}
for some mapping $\pi:\{1,...,K\}\times\{1,...,\alpha\}\rightarrow\{1,...,K\}$ and $\bd_{ij}\in\mathbb{R}^{\alpha}$.

Now, first, let us prove $\pi$ is only respect to $i$.
From Equation~\ref{eq:hypS1}, we know that for any $j\neq k$, $A(\a^{ij}+\a^{ik}) = A\a^{ij}+A\a^{ik} = B\bb^{ij}+B\bb^{ik} = B(\bb^{ij} + \bb^{ik})$.
Since $\a^{ij}+\a^{ik}$ is $1$-block-sparse, Equation~\ref{eq:hyp} implies that $\bb^{ij} + \bb^{ik}$ should also be $1$-block-sparse, which implies that $\pi(i,j) = \pi(i, k)$.
Hence, it follows that $\pi: \{1,...,K\}\rightarrow\{1,...,K\}$, and
\begin{equation}
A(e^K_i\otimes e^\alpha_j) = B(e^K_{\pi(i)}\otimes I_\alpha)\bd_{i j}.
\label{eq:hypS2}
\end{equation}

Then, let us prove that $D_i = [\bd_{i1},\dots,\bd_{i\alpha}]$ is invertible.
Denote $\a^i = [\a^{i1},\dots,\a^{i\alpha}]$ and $\bb^i = [\bb^{i1},\dots,\bb^{i\alpha}].$
From Equation~\ref{eq:hypS2}, it follows that $A\a^i = A[\a^{i1},\dots,\a^{i\alpha}] = A[(e^K_i\otimes e^\alpha_1),...,(e^K_i\otimes e^\alpha_\alpha)] = A(e^K_i\otimes I_\alpha),$
and $A\a^i = B\bb^i = B(e^K_{\pi(i)}\otimes I_\alpha)\begin{bmatrix}\bd_{i1},...,\bd_{i\alpha}\end{bmatrix} = B(e^K_{\pi(i)}\otimes I_\alpha)D_i$.
Therefore,
\begin{equation}
A(e^K_i\otimes I_\alpha) = B(e^K_{\pi(i)}\otimes I_\alpha)D_i.
\label{eq:hypS3}
\end{equation}
$A$ satisfies block RIP condition, and so $\text{rank}(A(e^K_i\otimes I_\alpha)) = \alpha$.
Form Equation~\ref{eq:hypS3}, $\text{rank}( B(e^K_{\pi(i)}\otimes I_\alpha)D_i) = \alpha$.
From basic linear algebra knowledge, we know $\text{rank}(XY) \le \min(\text{rank}(X), \text{rank}(Y)),$ for any matrix $X, Y$.
So $\text{rank}(D_i) \ge \alpha$.  
Since $D_i\in\mathbb{R}^{\alpha\times\alpha}$, $\text{rank}(D_i) = \alpha.$

Now, let us show $\pi$ is necessarily injective.
Suppose $\pi(i) = \pi(j)$, with $i\neq j$, then from Equation~\ref{eq:hypS3}, $A(e^K_i\otimes I_\alpha) = B(e^K_{\pi(i)}\otimes I_\alpha)D_i = B(e^K_{\pi(j)}\otimes I_\alpha)D_jD_j^{-1}D_i = A(e^K_j\otimes I_\alpha)D_j^{-1}D_i.$
Since $A$ satisfies RIP, which implies $A$ cannot map 2 different $s$-block-sparse vector to the same measurement, this is only possible if $i = j$. Thus, $\pi$ is injective.

Denote $P_\pi$ and $D$ as:
\begin{equation}
P_\pi = \begin{bmatrix} e^K_{\pi(1)} & \dots & e^K_{\pi(K)}\end{bmatrix}, D = \begin{bmatrix}D_1&\cdots & 0 \\ \vdots & \ddots & \vdots \\ 0 & \cdots & D_K\end{bmatrix}.
\end{equation}
Since $\pi$ is injective, $P_\pi$ is a permutation matrix.
Let us stack Equation~\ref{eq:hypS3} from left-to-right on both sides, and it follows that on left sides, $[A(e^K_1\otimes I_\alpha), \dots, A(e^K_K\otimes I_\alpha)] = A$, and on right sides, $[B(e^K_{\pi(1)}\otimes I_\alpha)D_1, \dots, B(e^K_{\pi(K)}\otimes I_\alpha)d_K] = B(P_\pi \otimes I_\alpha)D$.
Hence, finally, we proved in simple case $A = B(P_\pi\otimes I_\alpha)D.$

Before proving block ACS theorem, we first prove a proposition, and then use it for proving block ACS.

We use the same notation reported in ~\cite{hillar2011ramsey}: Denote $[K]$ for the set $\{1,\dots, K\}$ and $\begin{pmatrix}[K]\\s\end{pmatrix}$ for the set of $s$-element subset of $[K]$.
Moreover, denote the dictionary $A = [A_1, \dots, A_K]$ with $A_i\in\mathbb{R}^{P\times\alpha}$, and $\Span\{A_\mS\} = \{\sum_{i\in\mS} t_iA_i\}$.

{\it Proposition:}
Suppose that $A$ satisfies block RIP condition and that 
$$\kappa: \begin{pmatrix} [K] \\ s \end{pmatrix} \rightarrow \begin{pmatrix} [K] \\ s \end{pmatrix}$$
is a mapping with the following property: for all $\mS\in\begin{pmatrix} [K] \\ s \end{pmatrix}$,
\begin{equation}
\Span\{A_\mS\} = \Span\{B_{\kappa(\mS)}\}.
\label{eq:mapping1}
\end{equation}
Then, there exist a permutation matrix $P_\kappa\in\mathbb{R}^{K\times K}$ and an invertible block diagonal matrix $D \in \mathbb{R}^{\alpha K\times \alpha K}$ such that $A = B(P_\kappa\otimes I_\alpha)D$.

\begin{proof}
We prove from $s$ to $1$ inductively, and the final case $s = 1$ having already been worked out at the simple case.

First, let's show function $\kappa$ is injective.
Suppose that $\mS, \mS' \in \begin{pmatrix} [K] \\ s \end{pmatrix}$ are different and $\kappa(\mS) = \kappa(\mS')$ holds, then by Equation~\ref{eq:mapping1}, $\Span\{A_{\mS}\} = \Span\{B_{\kappa(\mS)}\} = \Span\{B_{\kappa(\mS')}\} = \Span\{A_{\mS'}\}.$
From below {\it Lemma 1}, it turns out that $\mS = \mS',$ which implies $\kappa$ is injective.

Now, let us prove the {\it Proposition} inductively.
Denote $\iota = \kappa^{-1}$ as the inverse of $\kappa$.  Fix $\mS = \{i_1, ..., i_{s-1}\}\in \begin{pmatrix} [K] \\ s-1 \end{pmatrix}$,  and set $\mS_1 = \mS\cup\{p\}$ and $\mS_2 = \mS\cup\{q\}$ for some fixed $p, q\not\in \S$ with $p \neq q$.  (Since $s < K$, $K-(s-1) > 1$, thus, it's possible to find such $p$ and $q$.) From Equation~\ref{eq:mapping1}, we obtain:
\begin{equation}
\Span\{A_{\iota(\mS_1)}\} = \Span\{B_{\mS_1}\},
\label{eq:S1}
\end{equation}
\begin{equation}
\Span\{A_{\iota(\mS_2)}\} = \Span\{B_{\mS_2}\}.
\label{eq:S2}
\end{equation}
Let us intersect Equation~\ref{eq:S1} and Equation~\ref{eq:S2}, and from below {\it Lemma 2} it follows that $\Span\{B_{\mS_1}\} \cap \Span\{B_{\mS_2}\} = \Span\{A_{\iota(\mS_1)\cap\iota(\mS_2)}\}$.
Since $\Span\{B_\mS\} \subseteq \Span\{B_{\mS_1}\} \cap \Span\{B_{\mS_2}\}$, it follows that $\Span\{B_\mS\} \subseteq \Span\{A_{\iota(\mS_1)\cap\iota(\mS_2)}\}.$
The number of the elements in $\iota(\mS_1)\cap\iota(\mS_2)$ is $s-1$, since $\iota(p) \neq \iota(q)$, with $p\neq q$, by injectivity of $\iota$.  Moreover the number of the elements in $\mS$ is also $s-1$, which implies that
\begin{equation}
\Span\{B_\mS\} = \Span\{A_{\iota(\mS_1)\cap\iota(\mS_2)}\}.
\label{eq:equal span}
\end{equation}
The association $\mS\rightarrow \iota(\mS_1)\cap\iota(\mS_2)$ from Equation~\ref{eq:equal span} defines a function $\sigma: \begin{pmatrix} [K] \\ s-1 \end{pmatrix} \rightarrow \begin{pmatrix} [K] \\ s-1 \end{pmatrix}$, with property that $\Span\{B_\mS\} = \Span\{A_{\sigma(\mS)}\}$.

Finally, let's show that $\sigma$ is injective. 
Suppose $\mS, \mS' \in \begin{pmatrix} [K] \\ s-1 \end{pmatrix}$, and $\sigma(\mS) = \sigma(\mS')$,  it follows that $\Span\{B_\mS\} = \Span\{A_{\sigma(\mS)}\} = \Span\{A_{\sigma(\mS')}\} = \Span\{B_{\mS'}\}$, from below {\it Lamma 1}, it follows that $\mS = \mS'$, which implies $\sigma$ is injective.
Hence, let $\xi = \sigma^{-1},$ with properties: for all $\mS\in\begin{pmatrix}[K]\\s-1\end{pmatrix}$, $\Span\{A_\mS\} = \Span\{B_{\xi(\mS)}\}.$
\end{proof}

{\it Lemma 1:}
If the dictionay $A = [A_1, \dots, A_K]$ satisfies block $2s$-RIP with RIP constant $\delta<1$, then for $\mS, \mS' \in  \begin{pmatrix} [K] \\ s \end{pmatrix},$
\begin{equation}
\Span\{A_\mS\} = \Span\{A_{\mS'}\}\hspace{3 mm}\Rightarrow\hspace{3 mm} \mS=\mS'.
\end{equation}

\begin{proof}
Suppose that $\mS\neq \mS'\in\begin{pmatrix} [K] \\ s \end{pmatrix}$ satisfying $\Span\{A_\mS\} = \Span\{A_{\mS'}\}$.  Then without loss of generality, there is an $i\in \mS$ with $i \not\in \mS'$, but atoms $A_i \in \Span\{A_{\mS'}\}$, which implies that the RIP constant $\delta = 1$, a contradiction to the assumption on $A$.
\end{proof}

{\it Lemma 2:}
If the dictionay $A$ satisfies block $2s$-RIP with RIP constant $\delta<1$, then for $\mS, \mS' \in  \begin{pmatrix} [K] \\ s \end{pmatrix},$
\begin{equation}
\Span\{A_{\mS\cap \mS'}\} = \Span\{A_{\mS}\}\cap \Span\{A_{\mS}\}.
\end{equation}

\begin{proof}
The inclusion ``$\subseteq$'' is trivial, so Let us prove ``$\supseteq$''. Suppose a block vector $\y\in \Span\{A_{\mS}\}\cap \Span\{A_{\mS_2}\}$.  Express $\y$ as a linear combination of $s$ atoms of $A$ indexed by $\mS$ and, separately, as a combination of $s$ atoms of $A$ indexed by $\mS'$. By RIP condition, these linear combinations must be identical. In particular, $\y$ was expressed as a linear combination of atoms of $A$ indexed by $\mS\cap \mS'$, and thus is in $\Span\{A_{\mS\cap \mS'}\}$
\end{proof}

{\it Proof of block ACS Theorem:}
Fix $\mS = \{x_1, x_2, ..., x_s\} \in \begin{pmatrix} [K] \\ s \end{pmatrix}$ and express $\a = \sum_{i = 1}^s\sum_{j = 1}^\alpha t^i_j \left(e_{x_i}^K\otimes e_{j}^\alpha\right),$
where $t^i_j\in \mT^i_j,$ a finite subset of $\mathbb{R}$. 
Suppose that $\bb$ is $s$-block-sparse with $A\a = B\bb$, and then, $\bb\in \Span\{e_{p_i}^K\otimes e_{j}^\alpha\big| i = 1,...,s, j = 1,..., \alpha\}$, for some $\{p_1, p_2, ..., p_s\}$.
Viewing each $s$-element subset of $[K]$ as a {\it color}, this map
\begin{equation}
f:\mT^1_1\times\dots\times \mT^1_\alpha\times \mT^2_1\times \dots\times \mT^s_\alpha\rightarrow\begin{pmatrix} [K] \\ s \end{pmatrix}
\end{equation}
\begin{equation}
\ie (t^1_1,\dots, t^1_\alpha, t^2_1,\dots, t^s_\alpha)\rightarrow \{p_1, p_2, ..., p_s\}
\end{equation}
is a coloring of the finite set $\mT^1_1\times\dots\times \mT^s_\alpha$ with colors in $C = \begin{pmatrix} [K] \\ s \end{pmatrix}.$\\

From Ramsey theory~\cite{hillar2011ramsey}, it implies that there are 2-element subsets $\mH^i_j\subseteq \mT^i_j$, and $\{r_1, r_2, ..., r_k\}\in \begin{pmatrix} [K] \\ s \end{pmatrix},$ such that $f(t^1_1, ..., t^s_\alpha) = \{r_1, r_2,..., r_k\}$ holds for all $(t^1_1, ..., t^s_\alpha)\in \mH^1_1\times... \times \mH^s_\alpha$.
Now, let us define function
\begin{equation}
\kappa(\{x_1, x_2, ..., x_k\}) = \{r_1, r_2, ..., r_k\},
\label{eq:def a}
\end{equation}
where $\{r_1, r_2, ..., r_k\}$ is generated by above recipe.

{\it Claim:}
The mapping $\kappa$ defined by Definition~\ref{eq:def a} satisfies Equation~\ref{eq:mapping1}.

\begin{proof}
First let us prove
\begin{equation}
\Span\{A_\mS\} \subseteq \Span\{B_{Alpha(\mS)}\}.
\label{eq:in}
\end{equation}
Choose a pair of elements $h, h' \in \mH_1^1\times... \times \mH_{\alpha}^{s}$, which differ only in the $l$-th coordinate.
By construction, the vectors $Ah, Ah'$ are in the right-hand set of Equation~\ref{eq:in}.
Then, the difference $A(h-h')$ that is a nonzero scalar multiple of one of bases of the left-hand set of Equation~\ref{eq:in}, is also in the right-hand set of Equation~\ref{eq:in}.
Hence, every bases of the left-hand set of Equation~\ref{eq:in} is in the right-hand set of Equation~\ref{eq:in}, $\ie \Span\{A_\mS\} \subseteq \Span\{B_{\kappa(\mS)}\}$.

Moreover, the dimension of left-hand set is $sK$, and the dimension of right-hand set is at most $sK$.  Hence, $Span\{A_\mS\} = Span\{B_{\kappa(\mS)}\}.$
\end{proof}

Finally, from the {\it Proposition}, block ACS theorem is proven.

{\small
\bibliographystyle{ieee}
\bibliography{egbib}
}

\end{document}